\documentclass[aps,twocolumn,prb,superscriptaddress,floatfix]{revtex4-2}
\setcitestyle{super}
\usepackage{amsmath,mathtools,graphicx,xcolor,bm,multirow,enumitem,soul,subfigure,fontspec,setspace,xr,float,amsmath,mathtools,booktabs,graphicx,xcolor,bm,multirow,enumitem,soul,subfigure,fontspec,setspace,longtable,mhchem, amssymb}

\newcommand{\ST}[1]{{\color{black}{#1}}}

\allowdisplaybreaks

\def\>{\rangle} 
\def\<{\langle} 
\def\br{{\bf r}}
\def\bR{{\bf R}}
\def\b0{{\bf 0}}
\def\bk{{\bf k}}
\def\bq{{\bf q}}
\def\k{{\kappa}}
\def\a{{\alpha}}
\def\Rm{{\bR\hspace{0.5pt}m}}
\def\0m{{\b0\hspace{0.5pt}m}}
\def\aibte{{\textit{ai}BTE}}
\makeatletter
\def\ps@myrevtex{%
  \def\@oddhead{\hfill\thepage}%
  \def\@evenhead{\thepage\hfill}%
  \def\@oddfoot{}%
  \def\@evenfoot{}%
}
\makeatother

\def\utoden{Oden Institute for Computational Engineering and Sciences, The University of Texas at Austin, Austin, Texas 78712, USA}
\def\utphysics{Department of Physics, The University of Texas at Austin, Austin, Texas 78712, USA}
\begin{document}

\title{Meta-optimization of maximally-localized Wannier functions}

\author{Sabyasachi Tiwari}
\affiliation{\utoden}
\affiliation{\utphysics}

\author{Bruno Cucco}
\affiliation{\utoden}
\affiliation{\utphysics}

\author{{Viet-Anh Ha }}
\affiliation{\utoden}
\affiliation{\utphysics}

\author{Feliciano Giustino}
\email{fgiustino@oden.utexas.edu}
\affiliation{\utoden}
\affiliation{\utphysics}

\maketitle

\onecolumngrid
\vspace{-10pt}
\noindent\textbf{Maximally-localized Wannier functions are quantum wavefunctions resembling atomic orbitals that are used to describe electrons in condensed matter.\cite{marzari2012} Since their introduction in 1997,\cite{Marzari1997} these functions have become ubiquitous in \textit{ab initio} materials simulations, including applications in linear-scaling methods,\cite{Goedecker1999} strongly-correlated electron systems,\cite{Georges1996} quantum transport,\cite{Calzolari2004} electron-phonon interactions,\cite{Giustino2017} and topological materials.\cite{Gresch2017} Despite their widespread adoption in a vast software ecosystem,\cite{Marrazzo2024} Wannier functions have not yet attained their fullest potential {in the presence of entangled bands}, as their optimization remains challenging and labor-intensive. Here, we introduce a universal meta-optimization method that leverages \ST{workflow abstraction} and machine learning techniques like differential evolution and Bayesian optimization to generate globally optimized Wannier functions without human intervention. We demonstrate this approach through three applications: (i) autonomous interpolation of entangled band structures with millielectronvolt accuracy \ST{starting from coarse Brillouin zone grids}, (ii) thousand-fold acceleration of fully \textit{ab initio} Boltzmann transport calculations \ST{via the use of minimal coarse Brillouin zone grids}, and (iii) ultrafast high-throughput calculations of high-precision Wannier functions for large materials libraries. This \ST{work} brings calculations that previously required supercomputers within the reach of personal computers.
}
\vspace{10pt}
\twocolumngrid

\noindent\textbf{Introduction}\\
Wannier functions offer a localized representation of electrons in crystalline solids that is complementary to the standard Bloch representation.\cite{marzari2012} If we denote by $|n\bk\>$ an extended Bloch function with band index $n$ and Bloch wavevector $\bk$, Wannier functions are defined by the following generalized Fourier transform:
 \begin{equation}\label{eq:wannier-def}
   |\Rm\> = {\sum}_n\int\!\!\frac{d\bk}{V_{\rm BZ}} e^{-i\bk\cdot \bR}\, U_{nm\bk}\, |n\bk\>.
 \end{equation}
In this expression, the integral is performed over the crystal Brillouin zone (BZ), with volume $V_{\rm BZ}$, and $\bR$ denotes a vector of the crystal lattice. The unitary matrix $U_{nm\bk}$ is arbitrary; by suitably choosing this matrix, one can obtain Wannier functions $|\Rm\>$ that are as localized as possible, known as maximally-localized Wannier functions (MLWFs).\cite{Marzari1997,Souza2001} The spatial localization of MLWFs forms the basis for linear-scaling electronic structure methods, whereby quantum-mechanical operators are conveniently represented as sparse matrices.\cite{Ordejon1996,Skylaris2005} 

MLWFs are primarily employed in Wannier-Fourier interpolation. This technique involves generating one-body operators in the Bloch representation from their sparse Wannier representation by inverting Eq.~\eqref{eq:wannier-def}.\cite{Souza2001} In this approach, Bloch functions and associated operators are initially computed on a coarse BZ grid (BZ$_{\rm c}$, for instance 10$\times$10$\times$10 $\mathbf{k}$ vectors). All relevant quantities are then transformed into the MLWF representation via Eq.~\eqref{eq:wannier-def}. From this real-space representation, the inverse transform is used to obtain Bloch operators on a fine BZ grid (BZ$_{\rm f}$, for instance 100$\times$100$\times$100 $\mathbf{k}$ vectors), in complete analogy with Fourier interpolation with zero padding: 
  \begin{equation}\label{eq:interp}
    |n\bk\>, \, \bk\in {\rm BZ}_{\rm c} \xrightarrow{\hspace{0.5cm}}
    |\Rm\> \xrightarrow{\hspace{0.5cm}} |n\bk^\prime\>, \, \bk^\prime\in {\rm BZ}_{\rm f}.
  \end{equation}
This operation achieves a reduction in computational cost by several orders of magnitude, all the while ensuring minimal interpolation error. As a result, calculations of materials properties that were previously impossible have become routine; for example, much of the  recent progress in electron-phonon physics has been enabled by the adoption of MLWFs;\cite{Giustino2007,Giustino2017} similarly, the extensive data on topological materials available today was generated using MLWFs.\cite{Zhang2019t,Vergniory2019} During the past two decades, MLWFs have been incorporated into all major density-functional theory (DFT) packages,\cite{VASP, DFT2, Abinit, Pyscf, fleur, GPAW, Siesta, Wien2k} as well as many-body electronic structure codes,\cite{Yambo, EPW2, NanoTCAD, Woptic, Z2pack, Perturbo, Gollum} and have been used in thousands of scientific studies. 

The procedure to generate MLWFs, commonly referred to as Wannierization, involves determining a set of unitary matrices $U_{nm\bk}$ in Eq.~\eqref{eq:wannier-def} that minimize the spatial extent of each Wannier function $|\Rm\>$. For a set of bands that are separated from all other bands by finite energy gaps, called a ``composite'' manifold, the matrices $U$ are square matrices with rank equal to the size of the manifold (subscripts omitted for brevity); this scenario is typical of valence electrons in semiconductors and insulators. Optimization is achieved by minimizing the Berry-phase spread functional,\cite{Marzari1997} which is defined as $\Omega = \sum_m \<\0m|\br^2|\0m\> - \<\0m|\br|\0m\>^2$, where $\br$ is the electron position and $\<\cdots\>$ denotes the integral over the crystal. Minimization is performed via gradient descent, with a starting point provided by the initial projections $U_{nm\bk}^0 = \<n\bk|\phi_m\>$; the $|\phi_m\>$ represent localized trial functions, for example atomic orbitals.  

In the case of metals, and for conduction bands of semiconductors and insulators, composite sets of bands cannot be identified (except in some rare cases); instead, one finds a continuous spectrum of entangled bands. The Wannierization of entangled bands involves two steps: first, identifying a subset of bands that best approximates a composite manifold, a procedure referred to as ``disentanglement''; second minimizing the spread functional within this manifold, as for composite bands.\cite{Souza2001} Since in these cases there typically are more Bloch functions than Wannier functions, the matrices $U$ are rectangular. 
% In practice there is a rectangular mask and a square unitary matrix, but we want to keep the discussion at an abstract level

Despite significant progress in Wannierization,\cite{Damle2017, Vitale2020, Qiao2023} all current strategies face difficulties due to the presence of numerous local minima in the $\Omega(U)$ landscape. As a result, Wannierization often requires significant human trial-and-error, especially for systems with highly entangled bands, complex unit cells, or topological obstructions.\cite{Brouder2007} While several promising approaches have been developed to automate this process and reduce reliance on chemical intuition,\cite{Damle2017, Vitale2020, Qiao2023} obtaining high-quality MLWFs remains a labor-intensive task.\cite{Ponce2021} \\

\noindent\textbf{Results}\\
\noindent\textit{\textbf{Meta-optimization of MLWFs}}\\
To systematically address this challenge, we make the key observation that Wannierization can be viewed as a functional mapping $\mathcal{W}$ from the space of all possible trial matrices $U^0$ to the optimized matrix $U$, which locally minimizes the MLWFs spread functional near $U^0$:
  \begin{equation}\label{eq:map}
    \mathcal{W}\!:\, U^0 \xmapsto{\hspace{5pt}\mathcal{W}\hspace{5pt}} \mathcal{W}(U^0) = \mathrm{arg\,\,min}_{U} \,\Omega(U)\big|_{U\sim U^0}\,.
  \end{equation}
This viewpoint suggests that, by optimizing $\mathcal{W}$ over the trial matrices $U^0$, one could generate improved MLWFs. This corresponds to extending the local optimization into a global search over trial matrices, a process we refer to as meta-optimization.  

Meta-optimization naturally allows for the incorporation of additional criteria beyond those captured by $\mathcal{W}$ itself. For instance, one could require that the Wannier-interpolated band structures are as accurate as possible, enforce a specific number of MLWFs, or ensure that the MLWFs be real-valued, as we demonstrate below. To achieve this, we introduce the loss function:
  \begin{equation}\label{eq:loss_full}
    \mathcal{L}(U^0) = {\sum}_{i}\alpha_i\,\zeta_i[\mathcal{W}(U^0)] +  
    {\sum}_{{j}}\lambda_{{j}}\,\theta\!\left\{{N_j}-\eta_j[\mathcal{W}(U^0)]\right\}.  
  \end{equation}
The first term in this expression includes the features $\zeta_i$ to be optimized; for instance, $\zeta_i$ might represent quantities such as spatial spread or band interpolation error. The relative importance of these features is controlled by the \ST{dimensionless} weights $\alpha_i$ in the loss function. The second term in Eq.~\eqref{eq:loss_full} imposes constraints on the features $\eta_j$ by introducing \ST{dimensionless} penalties $\lambda_j$ whenever a feature exceeds the specified bound $N_j$. Here, $\theta$ is the Heaviside step function. For example, this term might be used to enforce that the MLWFs spread remains within a desired limit.
\ST{Usually, $\lambda_j \gg \alpha_i$ ensures that such undesired solutions are not considered viable during the optimization.}
%\hl{Need to move this technical statement to SI methods: throughout this manuscript, we use a constant value $\lambda_i=10^6$}

The minimization of the loss function $\mathcal{L}$ poses three key challenges: (i) The mapping $\mathcal{W}$ is not explicitly defined as a single function; rather, it consists of a series of independent calculations, using both a DFT code and a Wannierization code\cite{Wannier} to generate initial projections and related quantities. Furthermore, if the loss $\mathcal{L}$ includes the band interpolation error among its features, evaluating this error requires separate DFT calculations to establish the ground truth. (ii) The parameter space of $\mathcal{W}$ is {highly complex}, as the trial matrices $U_0$ are determined by both continuous variables (e.g., energy disentanglement window) and categorical labels (e.g., the selection of trial atomic orbitals). (iii) Standard gradient-based optimization of $\mathcal{L}$ is not feasible, as its gradients are either computationally prohibitive (e.g., for the band interpolation error) or non-existent (e.g., with respect to categorical labels).

\begin{figure*}
  \begin{center}
    \includegraphics[width=0.55\textwidth]{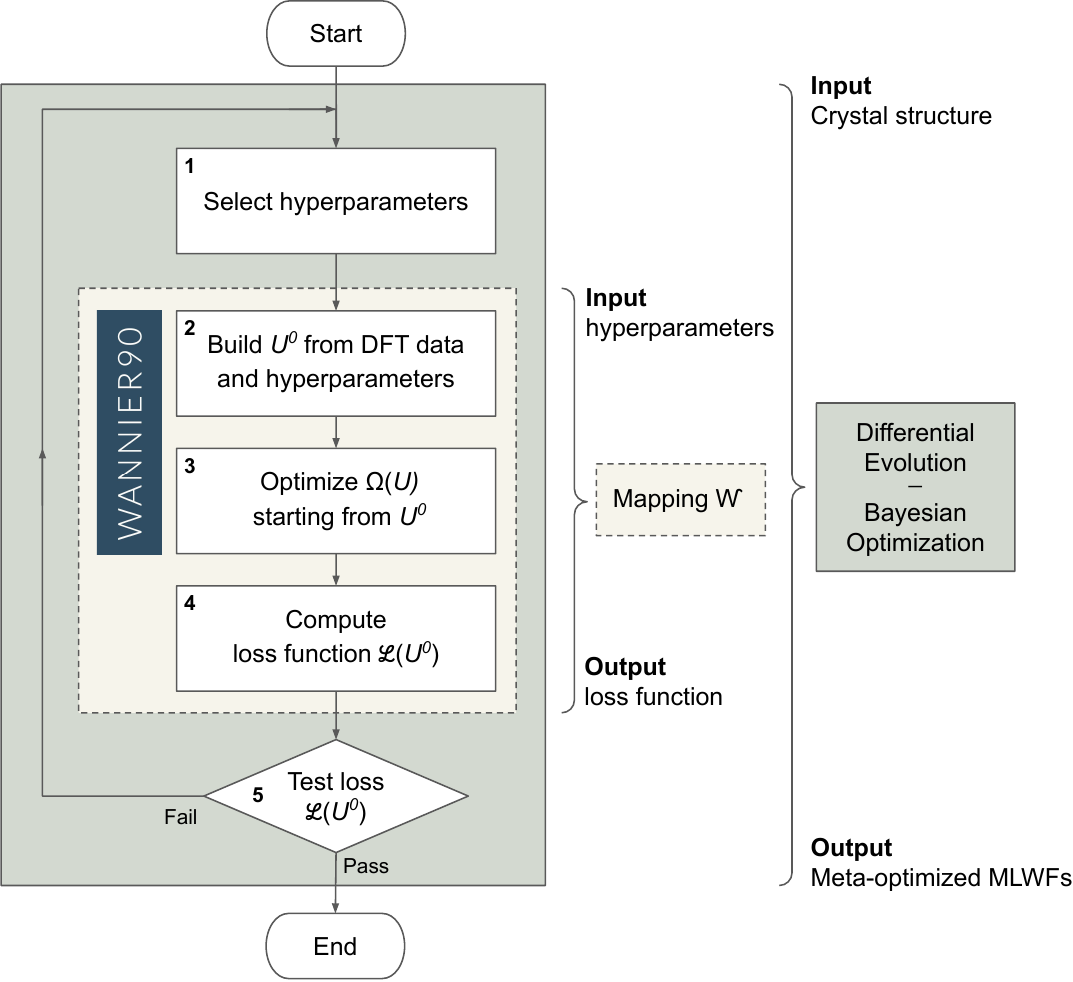}
    \caption{%
      % Link for figure:
      % https://docs.google.com/presentation/d/1wD-BLedRK3eejTKfyeajCtimgvFsRLGYxXIhtqE5Ge8/edit?usp=sharing
      %
      \ST{\textbf{Meta-optimization of Wannier functions via global optimization}. Schematic flowchart of the present method. The procedure starts with a standard DFT band structure calculation for a given crystal (in this work \texttt{Quantum ESPRESSO}\cite{DFT2} is used). At Step 1, the DE or BO algorithm selects the hyperparameters needed to construct $U^0$. In this manuscript, the hyperparameters are the energies of the inner and outer disentanglement windows and the type of initial projections, which are required in input by the \texttt{Wannier90} code.\cite{Wannier} Steps 2:4 implement the mapping $U^0 \mapsto \mathcal{W}(U^0)$: in Step 2, the hyperparameters are used to build $U^0$ using data from the DFT calculation; in Step 3, local optimization of the Berry-phase spread functional\cite{marzari2012} is performed using \texttt{Wannier90}; in Step 4, the loss function is computed. This step uses MLWFs obtained at Step 3, as well as band structures from separate DFT calculations for the ground truth. In Step 5, the loss function is tested; if the test fails, the procedure is repeated for a new DE generation or BO iteration. The mapping $\mathcal{W}$ (yellow box) is encapsulated within a single Python function, and is optimized by calling the procedure \texttt{differential\_evolution} of the SciPy library
      or \texttt{BayesianOptimization} of the \texttt{bayes\_opt} package
      (green box). The complete workflow is implemented in the \texttt{metaWann} code provided in code ocean capsule.~\cite{codeocean_capsule}
    }}\label{fig1}
  \end{center}
\end{figure*}

To address challenge (i), we employ \ST{workflow} abstraction, as shown Fig.~\ref{fig1}. The self-consistent DFT calculation, the non-self-consistent DFT step, and the Wannierization process are all encapsulated within a single function that encodes the mapping $\mathcal{W}$. Similarly, Wannier band interpolation and DFT band structure calculations, which are necessary for evaluating the band interpolation error (see \ST{Methods}), are encapsulated into a single abstract function. 

To address challenges (ii) and (iii), we employ machine-learning-inspired global optimizers that operate without relying on gradients, such as Differential Evolution (DE)\cite{Price1996,Storn1997} and Bayesian Optimization (BO).\cite{Mockus1975,mockus2005bayesian} DE and BO are particularly effective at identifying the global minimum of highly non-linear and non-differentiable objective functions, even within complex, multidimensional landscapes with numerous local minima. In particular, DE is a population-based evolutionary algorithm that iteratively refines a set of candidate solutions. At each generation, new candidates are created by a mutation step, which involves adding the weighted difference between two randomly selected vectors to a third vector. This is followed by a crossover step, whereby the new and old vectors are randomly combined, and the fittest vector is selected.\cite{Storn1997} In this work, the DE vectors correspond to the hyperparameters that are used to generate trial matrices $U^0$ (for example, using the \texttt{Wannier90} software\cite{Wannier}). 

Being a meta-optimization approach, this method can leverage the strengths of state-of-the-art Wannierization techniques rather than replacing them; for example, our approach can seamlessly be used with the selected columns of the density matrix (SCDM) method~\cite{Damle2017, Vitale2020} or the projectability disentanglement method.\cite{Qiao2023} The method is illustrated schematically in Fig.~\ref{fig1}, and the associated code is provided as {a Code Ocean capsule.}~\cite{codeocean_capsule} The flexibility in the choice of loss function and meta-optimizer can be used to fine-tune the method; for example, in Supplementary Figure~1 we investigate several possible loss functions, and in Supplementary Figure~2 we compare the performance of DE and BO.\cite{Mockus1975,mockus2005bayesian}\\

\begin{figure*}
  \begin{center}
    \includegraphics[width=\textwidth]{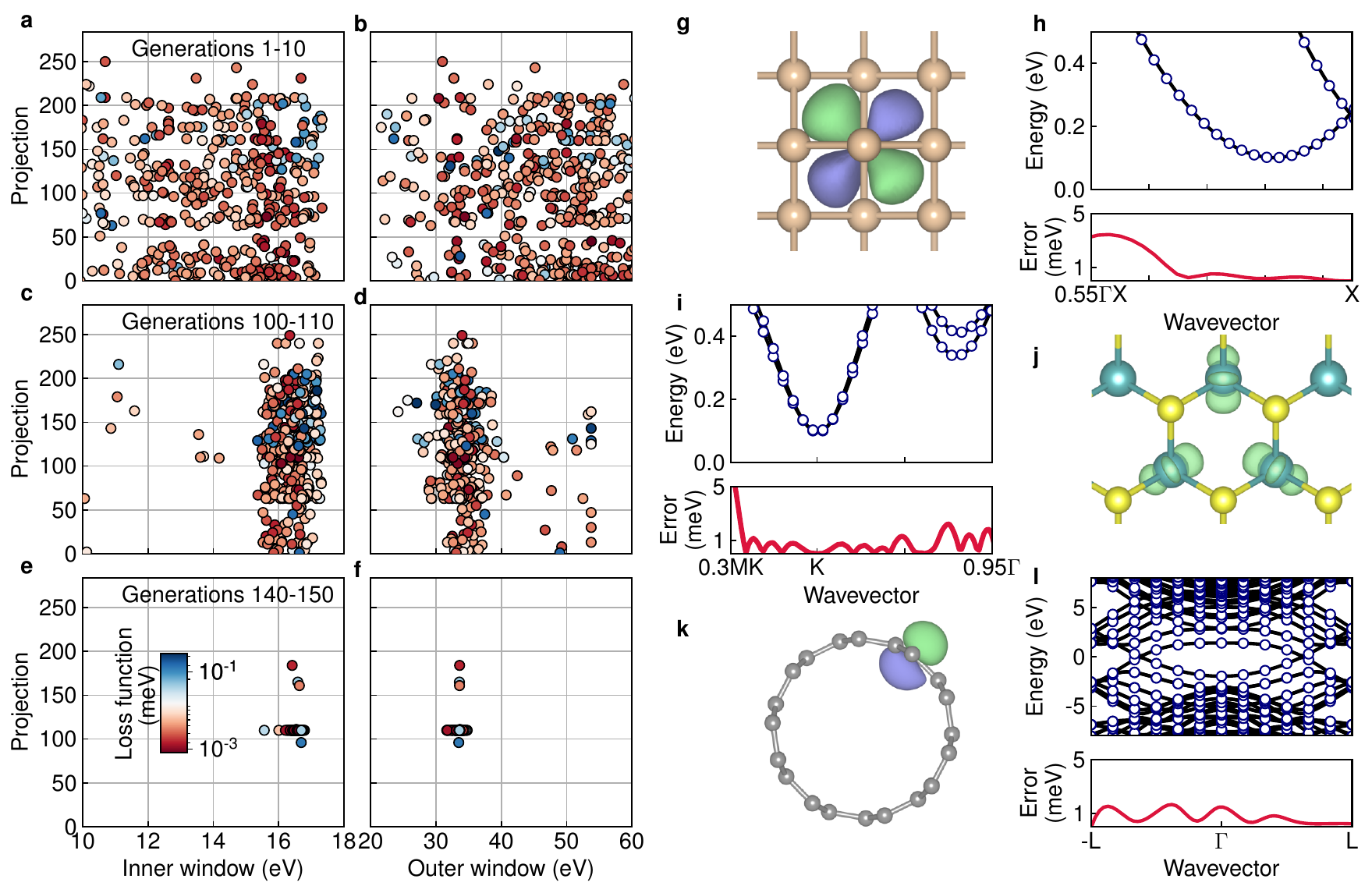}
    \caption{%
      \textbf{Wannier interpolation of entangled band structures with millielectronvolt accuracy}.  (a)-(f) 2D slices of the 5D hyper-parameter space explored by the DE meta-optimizer, for the conduction bands of silicon. The first, second, and third rows show data for DE generations 1:10, 100:110, and 140:150, respectively. Each disk corresponds to a single Wannierization, with red/blue indicating low/high loss function [colorbar in panel (e)]. We consider 285 types of initial projections, corresponding to hydrogenic orbitals with $s$, $p$, $d$, $f$ angular momenta, and a 6$\times$6$\times$6 coarse grid. We employ a DE population of 60 candidates, and evolve the system over 150 DE generations. (g) Meta-optimized MLWF of silicon, as obtained in (a)-(f). (h) Conduction bands of silicon: black lines are from Wannier interpolation using the MLWF shown in (g); disks are explicit DFT calculations. The underlying panel is the absolute value of the band interpolation error (red line). (i) Meta-optimized MLWFs for monolayer MoS$_2$: Comparison between DFT bands (disks) and Wannier-interpolated bands (black lines) near the conduction band bottom, and associated band interpolation error (red line).
      {Such a low interpolation error refers to an energy window within {300}~meV of the band bottom.} 
      A typical MLWF is shown in (j). Green and yellow atoms represent Mo and S, respectively. In this case we employ a 24$\times$24 coarse grid and 2 bands, a DE population of 30 candidates, and 50 DE generations. (k), (l) Meta-optimized MLWF of a (5,5) CNT and corresponding Wannier interpolation of band structures (disks: DFT, black lines: Wannier, red line: band interpolation error). In this case, the band interpolation error is evaluated over the two bands crossing the charge neutrality level (zero of the energy axis). For this case we employ a coarse one-dimensional BZ grid of 12 points, a DE population of 30 candidates, and 50 DE generations.
    }\label{fig2}
  \end{center}
\end{figure*}

\noindent\textit{\textbf{Benchmarks for interpolation accuracy}}\\
Figure~\ref{fig2} illustrates the present methodology for three prototypical materials across different dimensions: silicon in three dimensions, monolayer MoS$_2$ in two dimensions (2D), and a carbon nanotube (CNT) in one dimension. Figures~\ref{fig2}(a)-(f) show the optimization process for the conduction bands of silicon. Since the conduction band minimum does not reside at a high-symmetry point, achieving accurate Wannier interpolation is notoriously challenging in this case.\cite{Ponce2021} In the loss function, we include both the band interpolation error \ST{[Eq.~\eqref{eq:energyerror} of Methods]} and the spread, and we explore the space of initial matrices $U^0$ by varying the bounds of the inner and outer disentanglement energy windows as well as the type of initial projections.\cite{Wannier} Panels (a)-(b), (c)-(d), and (e)-(f) show two-dimensional cross-sections of this five-dimensional parameter space as meta-optimization progresses. Initially, candidate parameter vectors are nearly uniformly distributed; over generations, they tend to cluster near the global minimum. Concurrently, the band interpolation error decreases (see Supplementary Figure~1), until Wannier interpolation near the band {extrema} reaches sub-meV precision [Fig.~\ref{fig2}(h)]. This level of accuracy, starting from a coarse 6$\times$6$\times$6 BZ grid, is unprecedented. {This optimization identifies the $sp^3d^2$ hybrid orbital as the optimal initial projection, unlike the more common combination of $s$ and $d$ orbitals.}\cite{Ponce2021} {This result is consistent with the fact that silicon has both $sp^3$ and $d$ orbital character in the conduction.} Complete details on band interpolation are provided in Supplementary Figure~3. 

Figure~\ref{fig2}(i) illustrates the quality of band interpolation for monolayer MoS$_2$. This system features a multi-valley band structure, which poses significant challenges to Wannier interpolation. Using as little as two Bloch bands to generate Wannier functions, the present method achieves again sub-meV accuracy in band interpolation; importantly, this meta-optimization requires less than 30\,min on a personal computer (PC). Complete details on this test are reported in Supplementary Figure~4. 

Figures~\ref{fig2}(k)-(l) showcase the application of the present method to a (5,5) armchair CNT. Generating MLWFs for CNTs is notoriously difficult due to the large number of bands, the difficulty in identifying enough atomic orbitals for initial projections, and the presence of nearly-free electron states. To illustrate these challenges, Supplementary Figure~5 shows the result of manual Wannierization using carbon $p_z$ orbitals; this procedure leads to poor band interpolation {(when the $z$ axis does not follow the curvature of the tube)} that is not suitable for applications. In contrast, Fig.~\ref{fig2}(l) demonstrates that the present method achieves nearly perfect band interpolation, even for nearly-free electron states. For example, we achieve sub-meV precision throughout most of the BZ for the bands crossing the charge neutrality level. 

\ST{In Supplementary Figure~6} we report additional benchmarks for prototypical semiconductors, including AlSb, AlP, 3C-SiC, SrO, AlAs, and GaP. In all cases the present method achieves much higher interpolation accuracy than state-of-the-art manual Wannierization.~\cite{Ponce2021} \\

\begin{figure*}
  \begin{center}
    \includegraphics[width=1\textwidth]{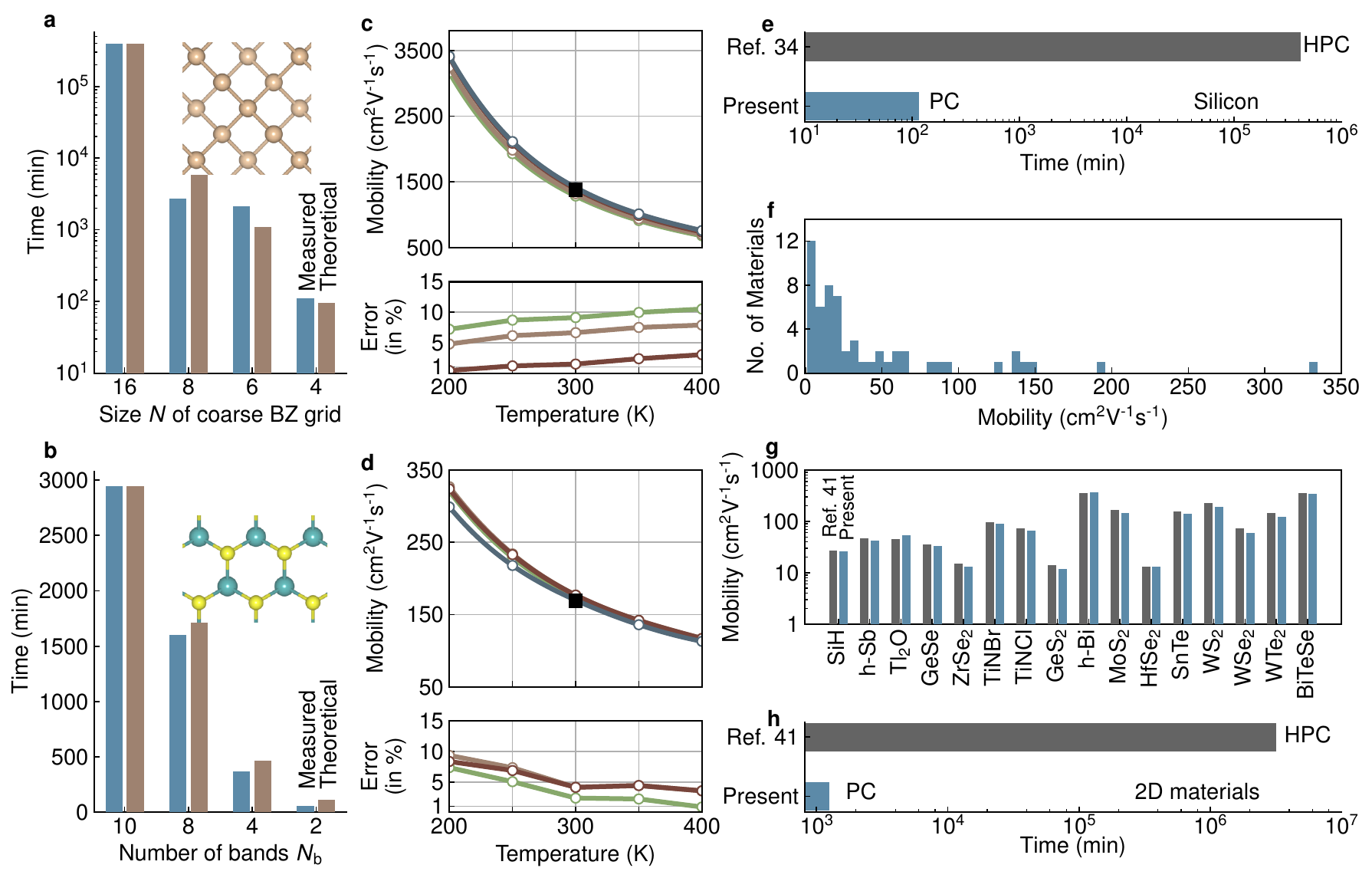} 
    \caption{
    \textbf{Acceleration of \aibte\ calculations of carrier transport using the evolutionary meta-optimizer}.  (a) Timing of \aibte\ calculations of the electron mobility in silicon, for different choices of the coarse BZ grid $N$$\times$$N$$\times$$N$; measured times (blue) are compared to ideal scaling time (brown).  The inset shows a ball-stick model of silicon. 
      (b) Total electron-phonon interpolation time of \aibte\ calculations of the electron mobility in monolayer MoS$_2$, for different choices of the number of Wannierized bands $N_{\rm b}$; measured times (blue) are compared to theoretical scaling times (brown), normalized to the values at $N_{\rm b}=10$. The inset shows a ball-stick model of monolayer MoS$_2$. The calculated mobilities from panels (a) and (b) are reported in Supplementary Table~2 for completeness.
      (c) Calculated temperature-dependent electron mobility of silicon, for different choices of the coarse BZ grid corresponding to the bar plot in panel (a), and relative error with respect to the densest grid ($N=16$). The linear grid size $N$ is indicated next to each curve. The corresponding band structures and Wannier functions are shown in Supplementary Figure~3. The square indicates the value from Ref.~\citenum{Ponce2021}. (d) Calculated temperature-dependent electron mobility of monolayer MoS$_2$, for different choices of the number of bands  corresponding to the bar plot in panel (b), and relative error with respect to the highest number of bands ($N_{\rm b}=10$). The corresponding band structures and Wannier functions are shown in Supplementary Figure~4. The square indicates the value from Ref.~\citenum{Ha2024}.
      (e) Speedup in calculations of the electron mobility of silicon. We compare the present runtime on a PC (48 Cascade lake cores, blue) with the runtime of Ref.~\citenum{Ponce2021} on the MareNostrum~4 supercomputer (with 480 Skylake cores, gray). Timing data is normalized to {a single} core in both cases. (f) Computed electron mobilities of 56 2D materials from the 2D Materials Cloud Database, shown as histogram. All values are reported in Supplementary Table~3, and the optimized hyperparameters are reported in Supplementary Table~1. (g) Comparison of mobilities of 2D materials calculated with the present method (blue) with state-of-the-art \aibte\ calculations for the 16 compounds reported in Ref.~\citenum{Ha2024} (gray). (h) Speedup in high-throughput calculations of the carrier mobility of 2D materials. The gray bar refers to the average runtime over the 16 compounds reported in (g), from Ref.~\citenum{Ha2024} (using the Frontera supercomputer, normalized to {a single} core). The blue bar is the average runtime over the 56 compounds in (f), using the present method on a PC.  
   }\label{fig3}
  \end{center}
\end{figure*}

\noindent\textit{\textbf{Acceleration enabled by meta-optimized MLWFs}}\\
\ST{As a second application, Fig.~\ref{fig3} demonstrates how the present method significantly accelerates carrier-mobility calculations in semiconductors based on the \textit{ab initio} Boltzmann transport equation (\aibte). The acceleration is achieved by using very small grids for the coarse BZ, and very few MLWFs, \textit{without} sacrificing accuracy.

It is well-known that the use of MLWFs drastically accelerates calculations of electron-phonon interactions via Wannier-Fourier interpolation of the electron-phonon matrix elements from a coarse BZ grid to an ultrafine grid.\cite{Giustino2007} The advantage of this method is that costly DFPT calculations are performed only on the coarse grid. 

However, even when employing Wannier-Fourier interpolation, \aibte\ calculations carry two bottlenecks: (i) the interpolation of electron-phonon coupling matrix elements on ultradense BZ grids, and (ii) the computation of the resulting carrier scattering rates. Wannier-Fourier interpolation of the electron-phonon matrix elements has a complexity scaling as $\mathcal{O}(N_{\rm c}^2)$, where $N_{\rm c}$ represents the size of the coarse BZ grid. For a coarse BZ grid with $N$ points in each direction, this leads to an overall scaling of $\mathcal{O}(N^6)$ for 3D materials. In addition to time complexity, the memory footprint is also substantial. The electron-phonon matrix elements in the Wannier representation form a five-dimensional array whose size scales as $N_{\rm c}^2$ and $N_{\rm b}^2$ ($N_{\rm b}$ is the number of Wannier functions). Consequently, reductions in $N_{\rm c}$ and $N_{\rm b}$ simultaneously decrease computational time and memory requirements. For example, a four-fold reduction in $N_{\rm c}$ theoretically yields a $4096$-fold speedup in transport calculations, while also significantly reducing storage requirements. Similarly, the computation of carrier scattering rates scales as $\mathcal{O}(N_{\rm b}^2)$; thus, a five-fold reduction in $N_{\rm b}$ can achieve an additional $25$-fold speedup. A detailed analysis of the scaling is provided in the Methods Section.}

% on 20 Dec we discussed with Sabya that the scaling should be:
% N_Rp N_q (N_b^2 + 2N_b^3) for ephwhan2blochp
% N_Re N_k N_b^2            for ephwan2bloch
% N_k and N_q are the fine grid vectors, and in BTE they are the same, so they don't matter and we have
%
% N_Rp (N_b^2 + 2N_b^3) for ephwhan2blochp
% N_Re N_b^2            for ephwan2bloch
%
% In the calculations for MoS2, we have N_Rp = 12^2, and N_Re = 24^2, so the relative scaling 
% of these two parts is:
%
% 12^2 * ( N_b^2 + 2N_b^3 ) for ephwhan2blochp
% 24^2 * N_b^2              for ephwan2bloch
%
% So for N_b>=2 ephwhan2blochp dominate, which should lead to Nb^3.
% However, in transport we use the fermi-window, therefore the indices m,n in (S3) do not scale
% with Nb but are fixed (only a subset of bands is computed). So the number of multiplications
% scales with the indices m',n' in (S2), which is O(Nb^2).
%
% We don't observe the N_b^2 scaling because the print_ibte (calculation of scattering rates 
% is a good fraction of the total time). But also the scattering rates don't scale as Nb^2 due 
% to the fermi-window. There, the scaling is dictaed by the fine grid.
%

We demonstrate these accelerations in Fig.~\ref{fig2}(a),(b) using silicon and  monolayer MoS$_2$. For silicon, we can solve the \aibte\ using a much smaller coarse BZ grid with 4$\times$4$\times$4 points, achieving a 3600-fold speedup compared to the standard 16$\times$16$\times$16 grid used in previous work,\cite{Ponce2021} with only a 8\% error [Fig.~\ref{fig3}(c)]. Similarly, for MoS$_2$, we can reduce the number of required MLWFs from 10 to 2 (see Supplementary Figure~4), achieving a 45-fold speedup as compared to previous work,\cite{Ha2024} with an error as low as 3\% [Fig.~\ref{fig3}(d)]. {The ability of the present method to achieve such a significant data compression is rationalized in Supplementary Figure~7 via a singular value decomposition of the matrix elements.}\cite{Luo2024}

As a result of the {speedup} enabled by meta-optimized MLWFs, calculations that once required large supercomputers can now be performed on standard PCs. To demonstrate this point, we repeated the state-of-the-art calculations of Ref.~\citenum{Ponce2021}, performed on the MareNostrum\,4 supercomputer, using a local PC. As shown in Fig.~\ref{fig3}(e), a complete \aibte\ calculation for silicon takes only 2 minutes and 20 seconds on the PC (with 48 Intel Cascade Lake cores) when using the present method, in contrast to 15 hours required on the supercomputer (with 480 Intel Skylake cores).\cite{PonceMaterialsCloud} \\\\

\noindent\textit{\textbf{Large-scale automation enabled by meta-optimized MLWFs}}\\
As a third application, we demonstrate the present method in autonomous high-throughput calculations of MLWFs and carrier mobilities for large materials libraries. Recently, the 2D Materials Cloud database~\cite{Talirz2020} was screened to identify 2D semiconductors with high charge carrier mobility.\cite{Ha2024} Out of 5,619 layered compounds, 258 candidate materials were found to be dynamically stable. Carrier mobilities were then computed using the \aibte\ for a reduced subset of 16 compounds with the lowest effective masses. Ideally, full-accuracy \aibte\ calculations would be needed for all 258 materials; however, generating reliable MLWFs for such a large library remains a significant challenge.

\ST{{Supplementary Figure~8} presents high-throughput calculations of MLWFs for a subset of 200 compounds from this library, showing that the present method achieves both predefined spatial localization and accurate band interpolation without any manual intervention.} Optimized hyperparameters, complete input files, and interpolated bands for all materials are {available on the Materials Cloud database}.~\cite{materialscloud2025} 
In Fig.~\ref{fig3}(f), we show the results of completely autonomous carrier mobility calculations via the \aibte\ for 56 of these compounds; in panel (g), we compare our results to the state-of-the-art calculations from Ref.~\citenum{Ha2024}. The agreement between our results and prior work is excellent; at the same time, our calculations are approximately 3000 times faster. In particular, we are able to perform transport calculations for all 56 compounds in the record time of 37 hours, i.e., with less than 40\,min per compound, on a local PC [Fig.~\ref{fig3}(h)]. This \ST{very significant acceleration opens the way to high-throughput screening} of carrier transport using the full \aibte. We emphasize, however, that detailed convergence tests on the coarse grid are always required for the most promising candidate materials.\\

\begin{figure*}[ht]
    \begin{center}
    \includegraphics[width=0.9\textwidth]{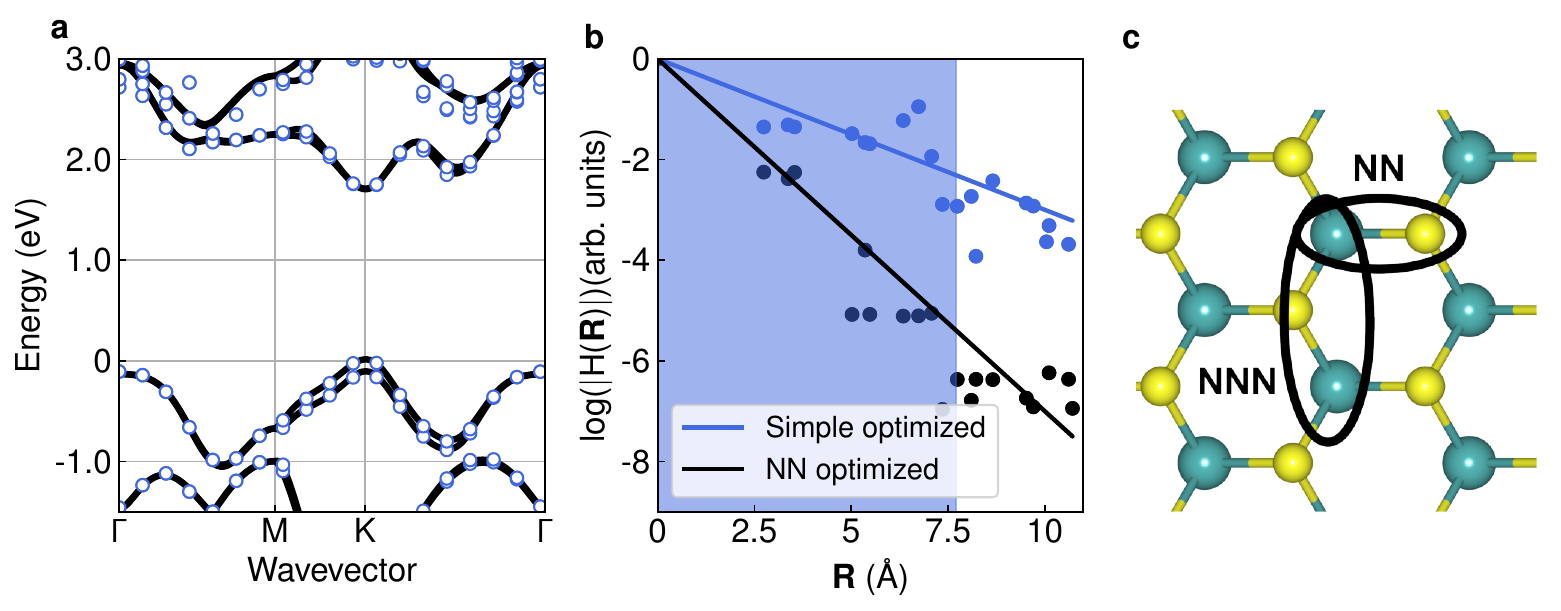}
    \caption{%
    \textbf{Building extreme tight-binding Hamiltonians with meta-optimization}.
    (a) Wannier-interpolation of the band structure of monolayer MoS$_2$ over a wide energy range. Black lines are results from the present method, blue disks are explicit DFT calculations. For this calculation, we employ an energy window in the loss function that extends from 0.8~eV below the valence band maximum to 0.5~eV above the conduction band minimum,  14 MLWFs, and a coarse BZ grid with 6$\times$6 points. The key difference between this calculation and those in Fig.\ST{2} and \ST{Fig.S4}~is that, here, we only retain the first and second nearest-neighbors in the Wannier representation. Specifically, in the interpolation of the smooth Bloch Hamiltonian starting from the Hamiltonian in the Wannier representation, $\tilde H_{mn}(\bk) = \sum_\bR \exp(i\bk\cdot\bR) H_{mn}(\bR)$, we only sum over $\bR=0$ and the first and second shell of lattice vectors $\bR\ne 0$. This truncated interpolation is used to evaluate the band interpolation error $\mathcal{E}$ in the loss function, and MLWFs are meta-optimized to minimize this error. (b) Spatial decay of the Hamiltonian in the Wannier representation,
    $|H_{mn}(\bR)|/\mathrm{max}[|H_{mn}(\bR)|]$. Blue disks correspond to the Hamiltonian obtained in Fig.~S4 (``simple optimized''), while black disks correspond to the present nearest-neighbor optimization (``NN optimized''). The lines are guides to the eye. The shading indicates the range of lattice vectors included in the nearest-neighbor optimization. We observe that the present optimization leads to a faster decay and improved Wannier interpolation. (c) Schematic illustration of the nearest-neighbors retained in (a) and (b).}\label{fig4}
    \end{center}
\end{figure*}

\begin{figure*}[ht]
    \begin{center}
        \includegraphics[width=0.90\textwidth]{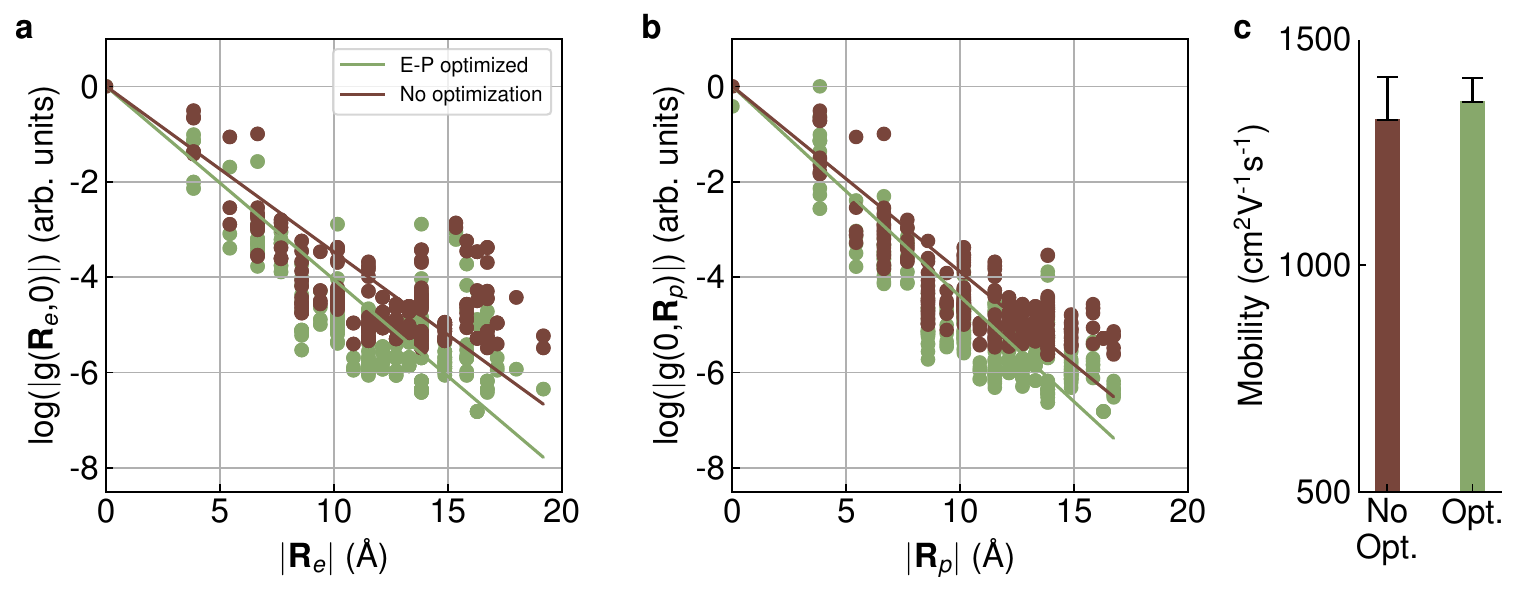}
        \caption{\textbf{{Maximally-localized electron-phonon matrix elements}}.
{Spatial decay of the electron-phonon matrix element for the conduction bands of silicon, as a function of (a) the length of the electronic Wigner-Seitz vector ($|{\bf R}_{e}|$) and (b) the length of the phonon Wigner-Seitz vector ($|{\bf R}_{p}|$). Brown disks correspond to calculation without including the term $\xi_2$ in the loss function, while green circles are obtained by including this feature. Solid lines are the corresponding exponential fits. (c) Electron mobility at 300\,K, obtained using electron-phonon-optimized Wannier functions (green) and standard Wannier functions (brown). The value on top of each bar is the relative error with respect to calculations using a dense coarse BZ grid.}}       
    \label{fig5}
    \end{center}
\end{figure*}

\noindent\textit{\textbf{Additional applications of meta-optimization}}\\
In addition to the \ST{acceleration of \aibte\ calculations demonstrated} above, meta-optimization can be customized for specific applications by designing more complex loss functions in Eq.~\eqref{eq:loss_full}, as discussed in the Supplementary Note~1. \ST{We demonstrate two such examples in this section. 

As our fourth example application, we use meta-optimization to construct a minimal, nearest-neighbor tight-binding Wannier Hamiltonian. To achieve this, in the interpolation of the smooth Bloch Hamiltonian starting from the Hamiltonian in the Wannier representation, $\tilde H_{mn}(\bk) = \sum_\bR \exp(i\bk\cdot\bR) H_{mn}(\bR)$, we only sum over $\bR=0$ and the first and second shell of lattice vectors $\bR\ne 0$ (shown in Fig.~\ref{fig4} (c). This truncated interpolation is used to evaluate the band interpolation error $\mathcal{E}$ in the loss function, and MLWFs are meta-optimized to minimize this error. 

In Fig.~\ref{fig4} we show the Wannier-interpolation of the band structure of monolayer MoS$_2$ over a wide energy range; black lines are results from the present method, blue disks are explicit DFT calculations. The key difference between this calculation and those in Fig.~\ref{fig2} and Supplementary Figure~4 is that, here, we only retain the first and second nearest-neighbors in the Wannier representation, as shown in Fig~\ref{fig4}. 

In Fig.~\ref{fig4}(b) we show the spatial decay of the Hamiltonian in the Wannier representation, $|H_{mn}(\bR)|/\mathrm{max}[|H_{mn}(\bR)|]$. Blue disks correspond to the Hamiltonian obtained in Fig.~\ref{fig2}, while black disks correspond to the present nearest-neighbor optimization. The lines are guides to the eye. The shading indicates the range of lattice vectors included in the nearest-neighbor optimization. We observe that the present optimization leads to a faster decay and improved Wannier interpolation. 

In the fifth example application, we generate Wannier functions that optimize the spatial localization of the electron-phonon matrix elements. We achieve this by incorporating an objective function for these matrix elements in the loss function. To this end we define the loss function L3 as follows:  
\begin{align}
     \mathcal{L}[U^0] = \alpha_1\,\xi_1[\mathcal{W}(U^0)] +  \alpha_2\xi_2[\mathcal{W}(U^0)] + \nonumber \\
     \lambda\,\theta\!\left\{{\Omega_0 }-\Omega[\mathcal{W}(U^0)]\right\}.
     \label{eq:L3}
\end{align}
Here, the first term $\xi_1 = \mathcal{E}/E_{\rm H}$ is the band interpolation error, as defined in the Supplementary Note~1; the second term $\xi_2 = \ell/a_0$ represents the spatial decay of the electron-phonon matrix elements, expressed via the decay length of the exponential fit in real space ($a_0$ is the Bohr radius); the third term $\theta\!\left\{{\Omega_0 }-\Omega\right\}$ is a constraint on the spread of the MLWFs.
This loss function guides the optimization in a direction where both the band interpolation error and the spatial decay of the electron-phonon matrix elements are optimized.

{We apply the loss function of Eq.}~(5) {to obtain Wannier functions for the conduction band of Si using a coarse BZ grid with $6^3$ points. In Fig.}~\ref{fig5}{(a),(b) we show that the spatial decay of the electron-phonon matrix elements improves using the loss function in Eq.}~(5){ as compared to optimization performed without this feature. Furthermore, in panel (c) we see that the mobility calculated using this additional feature is in better agreement with the most converged result obtained using a denser coarse BZ grid ($16^3$ points), the error being 3.6\% at 300\,K instead of 6.5\%. Therefore, the present approach also enables the calculation of maximally-localized electron-phonon matrix elements, yielding higher calculation accuracy.}

In the Supplementary Information we provide further examples of these capability. For example, in {Supplementary Figure~9} we demonstrate the generation of purely real-valued MLWFs; this capability not only further reduces computational complexity by an additional factor of two (on top of the acceleration demonstrated above), but also halves storage requirements and offers interpretative advantages over complex-valued MLWFs. 
{In Supplementary Note~2 and Supplementary Figure~10, we demonstrate the optimization of spread-balanced Wannier functions}.\cite{Fontana2021} {In Supplementary Figure~11, we illustrate how meta-optimization also improves Wannier interpolation for composite band manifolds. In Supplementary Figure~12, we show how this method improves the interpolation accuracy of the SCDM method.\cite{Damle2017, Vitale2020}} 
}\\

\noindent\textbf{Discussion}\\
\ST{We presented a meta-optimization framework to construct MLWFs that deliver superior band interpolation accuracy at reduced computational cost, and we demonstrated the application of this method in a number of diverse areas, from ultrafast transport calculations to generation of accurate nearest-neighbor Wannier Hamiltonians.}

The present method is general, composable, and future-proof by design. It is not tied to any specific optimization algorithm or codebase, and can be applied as a meta-optimization layer on top of any current or future electronic structure engine. In this sense, the present approach is \ST{not meant to replace any one Wannierization method in particular, but to augment all of them} by systematically identifying the optimal hyperparameters for a given application.

\ST{Moreover, the meta-optimization framework is designed as a modular layer that can operate on top of existing Wannierization tools, including Wannier90 and all its workflows. Since it generates hyper-parameters for maximal localization, it can be readily integrated within the wider Wannier ecosystem, for instance WannierBerri and WannierTools.~\cite{Tsirkin2021,WU2017}}

Additional applications can be envisioned in rapid data generation for machine learning and AI~\cite{Butler2018, Merchant2023} to overcome the dilemma of small data,\cite{Xu2023} sparse matrix representations for large scale quantum device simulations,\cite{Afzalian2021} and high accuracy downfolded Hamiltonians for strongly correlated systems,\cite{Chang2024,Arita2015} quantum algorithms,\cite{Alvertis2024} and second-principles simulations.\cite{Zubko2016} By realizing the full potential of MLWFs with meta-optimization, it will also become possible to perform advanced electronic structure calculations on standard PCs instead of large supercomputers, as demonstrated here in the context of electron transport.
\ST{Looking ahead, we anticipate that the integration of workflow abstraction with multi-objective meta-optimization will find broader applications in atomic-scale materials design.} \\

\noindent\textbf{Methods}\\
We here summarize the key aspects of the methodology. Details on the computational setup, such as software, parameters, and loss functions are provided in the Supplementary Information.\\[3pt]

\noindent\textit{\textbf{Loss function}}.
The loss function used for meta-optimization is given by Eq.~(4). The $\zeta_i$ represent objective functions to be optimized, such as for example the spread $\Omega$ of MLWFs or the band interpolation error. Similarly, the functions $\eta_j$ are constrained to the values $N_j$ by adding the penalties $\lambda_j$ to the loss function whenever $\eta_j$ exceeds the bound $N_j$ (as determined by the Heaviside step function $\theta$). We define the band interpolation error $\mathcal{E}$ as the root mean square deviation between DFT bands and Wannier-interpolated bands within a given energy window $[\mu_{\rm min},\mu_{\rm max}]$: 
  \begin{equation}\label{eq:energyerror}
    \mathcal{E}= \left[\frac{\displaystyle{\sum}_{n\bk}\,f^{\mu_{\rm max}}_{n\bk}(1-f^{\mu_{\rm min}}_{n\bk})\left(\varepsilon^{\rm DFT}_{n\bk}-\varepsilon^{\rm W}_{n\bk}\right)^2}{ \displaystyle  {\sum}_{n\bk}\,f^{\mu_{\rm max}}_{n\bk}(1-f^{\mu_{\rm min}}_{n\bk})}\right]^{1/2}\!\!\!. %\label{e:E}
  \end{equation}
In this expression, $\varepsilon_{n\bk}$ denotes the Kohn-Sham eigenvalue for band $n$ and wavevector $\bk$, the superscript ``DFT'' refers to explicit DFT calculations, and the superscript ``W'' refers to eigenvalues obtained by Wannier interpolation. We include a rectangular filter via the Fermi-Dirac functions $f^{\mu_{\rm min}}_{n\bk}$ and $f^{\mu_{\rm max}}_{n\bk}$, which are defined as
$f^\mu_{n\bk} = \{\exp[(\varepsilon_{n\bk}-\mu)/\sigma]+1\}^{-1}$; we use a broadening parameter $\sigma=1$~meV to avoid discontinuous jumps in the loss function. The sums are evaluated over all bands and all $\bk$ points included in the Wannierization process. 
The bounds $\mu_{\rm min}$ and $\mu_{\rm max}$ are chosen depending on the specific application. For example, accurate calculations of phonon-limited carrier mobilities requires highly-accurate band interpolation within a small energy window of the order of 300-500 meV near the Fermi level or band edges; while optical absorption calculations may  require an energy range in the multi-eV range around the Fermi level. 
In the Supplementary Figure~1, we investigate several possible loss functions. \\[3pt]
\noindent\textit{\textbf{Choice of meta-optimization algorithm}}.
Our method can utilize any global optimization algorithm since the meta-optimization step is independent from the Wannierization step, as shown in Fig~\ref{fig1}. 
However, we employ DE for meta-optimization for results shown in Fig.~\ref{fig2},~\ref{fig3} as well as for most of the applications.\cite{Storn1997} Our choice is motivated by the fact that DE is a robust global optimizer that does not rely on gradients. We expect that other metaheuristic optimization methods will produce similar results.\cite{Talbi2009} For completeness, in Supplementary Figure~2 we report a comparison between DE and Baysian optimization (BO) which is a standard tool in machine learning.\cite{Mockus1975,mockus2005bayesian} 
BO is a probabilistic optimization technique that uses a surrogate model, typically a Gaussian process, to guide the search for the optimum by balancing exploration and exploitation of the objective function. BO begins by randomly sampling the parameter space to evaluate the loss function. Using these initial evaluations, BO creates a probabilistic model of the loss function by defining a prior and then updating it to form a posterior distribution. From the posterior, an acquisition function is then constructed to guide the selection of the next set of parameters to evaluate. One potential disadvantage of BO is that, when the loss function is not smooth (as in the present case which involves categorical labels), BO may get stuck in a local minimum. Supplementary Figure~2 shows how DE offers superior performance than BO in the present problem of meta-optimization of MLWFs. For these calculations we employ the \texttt{bayesopt} function of the Bayesian-Optimization Python library~\cite{Bayesopt}. For both the DE and BO, we use the default hyperparamter settings as provided by the scipy~\cite{Scipy} and the Bayesian-Optimization~\cite{Bayesopt} libraries.\\[3pt]
\noindent\textit{\textbf{Scaling of electron-phonon calculations based on MLWFs: \aibte~as a representative example}}. 
\aibte\ calculations involve two time-consuming steps: (i) The evaluation of electron-phonon matrix elements on fine BZ grids using Wannier-Fourier interpolation, and (ii) the evaluation of the carrier scattering rates starting from these matrix elements and the electron and phonon band structures. Details of these calculations are provided in Refs.~\citenum{Ponce2018,Lee2023}. The evaluation of electron-phonon matrix elements on the fine BZ grid is performed via:\cite{Giustino2007}
\begin{align}
   g_{mn\nu}(\bk,\bq) & =  \sum_{pp'}
  e^{i(\bk\cdot \bR_p+\bq\cdot \bR_{p'})}
  \nonumber\\
    & \sum_{\,\,m'n'\k\a} U_{m m'\bk+\bq}\, 
    g_{m'n'\k\a}(\bR_p,\bR_{p'})\, 
    U^\dagger_{n'n\bk} {\rm u}_{\k\a,\bq\nu}. %\numberthis 
    \label{eq.g}
\end{align}
In this expression, $g_{m'n'\k\a}(\bR_p,\bR_{p'})$ are matrix elements in the Wannier representation, with $m',n'$ running over the number of Wannier functions $N_{\rm b}$, $\k$ representing the atom label, and $\alpha$ the Cartesian direction for the variation of the Kohn-Sham potential, and $\bR_p,\bR_{p'}$ the lattice vectors of the unit cells where Wannier functions are located. The unitary matrices
$U_{n'n\bk}$ are used to transform the Wannier matrix elements into the smooth Bloch representation for the electron wavevector $\bk$,\cite{Marzari1997} and the matrices ${\rm u}_{\k\a,\bq\nu}$ represent the atomic eigendisplacements for the phonon mode $\nu$ with wavevector $\bq$. The summations are over all atoms ($\kappa$), Cartesian directions ($\alpha$), Wannier functions ($m',n'$), and lattice vectors ($p,p'$). The resulting matrix element $g_{mn\nu}(\bk,\bq)$ represents the probability amplitude for an electron in state $|n\bk\>$ to scatter into the state $|m\bk+\bq\>$ via the phonon $\bq\nu$.
The scattering rate needed in the solution of the Boltzmann transport equation is given by:\cite{Ponce2018,Lee2023,Giustino2017}
\begin{align}\label{eq.scatt}
  \tau_{n\bk}^{-1} = & \frac{2\pi}{\hbar} \sum_{m\nu} \!\int\! \frac{d\bq}{\Omega_{\mathrm{BZ}}} 
  | g_{mn\nu}(\bk,\bq)|^2  \nonumber \\
  &\Big[ (n_{\bq\nu} +1 - f_{m\bk+\bq}^0) 
  \delta( \varepsilon_{n\bk} - \varepsilon_{m\bk+\bq}   -  \hbar \omega_{\bq\nu}) \nonumber \\
   & +   (n_{\bq\nu}  +   f_{m\bk+\bq}^0 )\delta(\varepsilon_{n\bk} - \varepsilon_{m\bk+\bq}  
   +  \hbar \omega_{\bq\nu}) \Big], %\numberthis
\end{align}
where $\hbar$ is the Planck constant, $n_{\bq\nu}$ is the Bose-Einstein occupation for the phonon $\bq\nu$, $f_{n\bk}^0$ is the Fermi-Dirac occupation for the electronic state $|n\bk\>$, and $\omega_{\bq\nu}$ is the phonon frequency. The functions $\delta(\cdots)$ are Dirac delta functions. 

From Eq.~\eqref{eq.g} we see that, if we keep the number of $\bk$ and $\bq$ points in the fine BZ grids as well as the number of Wannier functions fixed, the number of complex multiplications is proportional to the number of lattice points $\bR_p$ and $\bR_{p'}$ in the sums. These points belong to Wigner-Seitz cells whose dimensions are (nearly) equal to the number of $\bk$-points and $\bq$-points on the coarse BZ grids. In the case of equal $\bk$- and $\bq$-grids with $N_{\rm c}$ points, the total number of multiplications is proportional to $N_{\rm c}^2$. Furthemore, from Eq.~\eqref{eq.g} we also see that, for each $\bk$- and $\bq$-point and for each lattice point $\bR_p$ and $\bR_{p'}$, we need to evaluate $2N_{\rm b}^3$ products $U_{m m'\bk+\bq}\, g_{m'n'\k\a}(\bR_p,\bR_{p'})\, U^\dagger_{n'n\bk}$. Therefore, in the most general case, the scaling of the electron-phonon Wannier interpolation with the coarse grid size and with the number of Wannier functions is $\mathcal{O}(N_{\rm c}^2N_{\rm b}^3)$.\\
In the specific case of \aibte\ calculations, it is standard practice to only evaluate Bloch states within a predefined energy window centered at the Fermi level;\cite{Lee2023} this choice is justified by the fact that the scattering rates given by Eq.~\eqref{eq.scatt} are nonzero only in the vicinity of the Fermi level. With this choice, the indices $m,n$ in Eq.~\eqref{eq.g} are constrained by the energy window and do not scale with $N_{\rm b}$, therefore the evaluation of the matrix elements requires $\mathcal{O}(N_{\rm c}^2N_{\rm b}^2)$ products.

From Eq.~\eqref{eq.scatt} we see that, for each $\bk$-point in the fine grid and for each electronic band $n$, we need to evaluate the terms within the square brackets $N_{\rm b}$ times. Therefore this operation should have a theoretical scaling with respect to the number of Wannier functions of $\mathcal{O}(N_{\rm b}^2)$. However, as for the matrix elements, the use of an energy window centered at the Fermi level effectively makes the scaling of the scattering rates independent of $N_{\rm b}$. In this case, what dictates the compute time is the size of the fine BZ grid. 

The above discussion focuses on the \aibte\ case, but the same scaling considerations apply more broadly to calculations relying on Wannier-interpolated electron-phonon matrix elements, including phonon-assisted optics,~\cite{Tiwari2024,KP} phonon-mediated superconductivity,~\cite{Margine2013} and polarons.~\cite{Sio2019,Sio2023}

\vspace{10pt}
\noindent\textbf{Data Availability}\\
All data presented in this manuscript are reported {in the Supplementary Tables and in the Materials Cloud database.}~\cite{materialscloud2025}

\vspace{10pt}
\noindent\textbf{Code Availability}\\
The codes used in this work, namely \texttt{EPW}, \texttt{Quantum ESPRESSO}, \texttt{Wannier90}, and \texttt{EPWpy} are all open-source software and are freely available from their respective websites. For reproducibility purposes, the {\texttt{metawann} source code and an example are available and executable as a a Code Ocean capsule.}~\cite{codeocean_capsule}

\noindent\textbf{Acknowledgments}\\
This work was supported by the Computational Materials Science program of the U.S. Department of Energy, Office of Science, Basic Energy Sciences, through award no. DE-SC0020129. Computational resources were provided by the National Energy Research Scientific Computing Center (a DOE Office of Science User Facility supported under Contract No.~DE-AC02-05CH11231), the Argonne Leadership Computing Facility (a DOE Office of Science User Facility supported under Contract DE-AC02-06CH11357), and the Texas Advanced Computing Center (TACC) at The University of Texas at Austin. 

\vspace{10pt}
\noindent\textbf{Author Contributions}\\
S.T. and F.G. designed the project. S.T. wrote the \texttt{EPWpy} library and the {\texttt{metawann}} code. B.C. contributed to the development and testing of \texttt{EPWpy} and {\texttt{metawann}}. S.T, B.C., and {V.-A.H.} performed calculations and data analysis. F.G. supervised the project. All authors contributed to the writing of the manuscript.

\vspace{10pt}
\noindent\textbf{Competing Interests}\\
S.T., B.C., and F.G. are inventors on a pending U.S. provisional patent application (No. 63/730,258, filed 12/10/2024) entitled “Systems and methods for generating Wannier functions,” relating to the meta-optimization algorithm described in this work. The other authors declare no competing financial or non-financial interests.

\clearpage
\newpage

\bibliographystyle{naturemag}
\bibliography{references}

%\newpage

\newpage

\newpage

\newpage

\newpage

\clearpage

\setcounter{page}{1}
\pagenumbering{arabic}

\newpage

\onecolumngrid

\newcount\p
\p=1
\loop
  \clearpage
    \ifnum\p=1
    \thispagestyle{empty}   % no page number on first page
%  \else
%    \thispagestyle{plain}   % or keep default
  \fi
    \includegraphics[page=\the\p,width=1.0\textwidth]{Supplementary_cropped.pdf}
  \advance\p by 1
\ifnum\p<21
\repeat

\clearpage
\newpage

\setcounter{page}{1}
\pagenumbering{arabic}

\newcount\p
\p=1
\loop
  \clearpage
    \ifnum\p=1
    \thispagestyle{empty}   % no page number on first page
%  \else
 %   \thispagestyle{plain}   % or keep default
  \fi
    \includegraphics[page=\the\p,width=\textwidth]{Supplemental_File1.pdf}
  \advance\p by 1
\ifnum\p<607   % <-- change to number of pages
\repeat

\clearpage
\newpage

\end{document}